\documentclass[aps,twocolumn,preprintnumbers,floatfix,showpacs,prx]{revtex4-1}

\usepackage{amsmath,amssymb}
\usepackage{dcolumn}
\usepackage{braket}
\usepackage{dsfont}
\usepackage{feynmp}
\usepackage{hhline}
\usepackage{graphicx}
\usepackage{mathtools}
\usepackage{xfrac}
\usepackage{paralist}
\usepackage{enumitem}
\usepackage[normalem]{ulem}
\usepackage{color}
\definecolor{darkblue}{rgb}{0.,0.,0.4}
\definecolor{darkred}{rgb}{0.5,0.,0.}

\newcolumntype{L}{>{\centering\arraybackslash}m{12.4cm}}

\newcommand{\beq}{\begin{eqnarray}}
\newcommand{\eeq}{\end{eqnarray}}

\begin{document}

\title{Quantum Error Correction for Complex and Majorana Fermion Qubits}
\author{Sagar Vijay}
\author{Liang Fu}
\affiliation{Department of Physics, Massachusetts Institute of Technology,
Cambridge, MA 02139, USA}
\begin{abstract}
We introduce error-correcting codes that can correct for fermion parity-violating (quasiparticle poisoning) and parity-conserving errors in systems of complex fermions and of Majorana fermions.  After establishing properties of fermion codes, we introduce a generic construction of fermion codes from weakly self-dual classical, binary error-correcting codes.  We use this method to construct (i) the shortest fermion code to correct for quasiparticle poisoning errors, (ii)  translationally-invariant fermion codes and (iii) other codes that correct higher-weight errors.  We conclude by discussing a possible physical realizations of one such code in mesoscopic superconductors hosting Majorana zero modes.
\end{abstract}
\maketitle

Qubits for quantum information processing 
are mostly built from bosonic degrees of freedom, so that the two basis states of a physical qubit have the same fermion parity.
Recently, there has been a growing effort towards storing quantum information in fermion states, so that a qubit is encoded in the fermion occupation number ($n=0,1$) or equivalently, the fermion parity $(-1)^n=\pm 1$.  Fermion qubits may be constructed from localized sub-gap states in a superconductor \cite{Andreev, Yu,Shiba, Rusinov}. A remarkable recent experiment \cite{Urbina} has demonstrated the coherent manipulation of Andreev states in a superconducting weak link.

In addition to providing a new physical implementation of a quantum computer, using fermions as the carriers of quantum information necessitates a new computational model, as Fermi statistics forbids a mapping of local quantum gates acting on fermions to local gates acting on bosons \cite{BravyiKitaev}. For the same reason, quantum error correction -- which is essential for fault-tolerant quantum computation -- is different in systems of fermions and of bosons.
Quantum error correction in Fermi systems has received attention only recently \cite{Bravyi, Vijay, DiVincenzo, DiVincenzo_2}.

An important factor limiting the performance of quantum information processing with fermion qubits is quasiparticle poisoning, whereby a single electron tunnels between states that store information and unknown states in the environment, thereby changing the fermion parity of the encoded qubit, and resulting in an error. Various experiments indicate that the quasiparticle density in a gapped superconductor far exceeds its value in thermal equilibrium \cite{qp1,qp2}, and the origin of the excess quasiparticles is not well understood. Fermion parity flips due to quasiparticle poisoning have been directly observed in continuous, real-time measurements on spin-degenerate Andreev bound states in superconducting atomic contacts \cite{Urbina}. DC transport measurements on spin-polarized subgap states  in proximitized semiconductor nanowires have inferred possible parity lifetimes up to 10 ms \cite{Marcus-Parity, Marcus-Parity2}. To extend the lifetime of fermion qubits requires finding reliable ways of correcting both parity-conserving and parity-violating errors.

In addition to Andreev bound-states, spatially-separated zero-energy Majorana fermions (Majorana zero modes) \cite{Kitaev,Read-Green} have received recent attention as carriers of quantum information.  These fractional degrees of freedom are predicted to exist in topological superconductors. Following theoretical proposals, recent experiments have observed evidence for Majorana fermions in proximitized nanowires \cite{Copenhagen1, Copenhagen2, Delft}, atomic chains \cite{Princeton} and topological insulators \cite{Jia}. A variety of different approaches to quantum information processing with Majorana fermions have been proposed \cite{Kitaev,Bravyi,Vijay,VijaySurface2, Egger, Egger2,LossSpin,Vijay2,Box,StationQ}. The original work of Kitaev \cite{Kitaev} proposed storing one qubit in the fermion parity of a pair of spatially separated Majorana zero modes $\gamma_{1,2}$, which together form a complex fermion $f^\dagger= \gamma_1 + i\gamma_2$. However, unlike ordinary complex fermions, the parity operator in Majorana qubits ($i \gamma_1 \gamma_2$) is a non-local operator whose eigenvalue cannot be inferred from local measurements. It is thus expected that error processes involving multiple Majorana fermions are suppressed exponentially in their spatial separation. However, this separation does {\it not} offer protection against quasiparticle poisoning, which is a local process that involves individual Majorana fermions independently \cite{FuKane, Loss}.  Therefore, quasiparticle poisoning presents a serious challenge for the long-term prospects of quantum information processing with both complex and Majorana fermions.

Recent interest in error correction in Fermi systems has led to a new and more robust approach to Majorana-based quantum computation, in which qubits are encoded in a collection of Majorana fermions \cite{Vijay, Bravyi}; the added redundancy is advantageous for error correction. In Majorana fermion ``surface codes" \cite{Vijay, Egger}, measurements of commuting operators (``stabilizers") in a two-dimensional array of Majorana fermions are used for active error correction and measurement-based quantum computation.
A family of Majorana fermion codes has also been constructed from a restricted set of Pauli stabilizer codes for bosonic qubits \cite{Bravyi}.  In previous studies of fermion codes, however, fermion parity conservation has often been assumed and as a result, the issue of correcting fermion-parity violating errors  has not been addressed.





In this work, we introduce efficient fermion error-correcting codes that can correct for quasiparticle poisoning errors as well as other error processes.  These codes are inherently ``fermionic" in the sense that they cannot be mapped onto a bosonic code through local unitary transformations.   Error detection and correction, as in the stabilizer formalism \cite{Gottesman_Thesis}, is implemented through the projective measurement of a set of commuting multi-fermion operators, or stabilizers, that do not disturb the encoded qubit.  An essential feature of our codes that enables error correction is that the encoded qubit can only be determined by measuring a multi-fermion operator, in contrast to the standard encoding of a qubit in a fermion bilinear \cite{Kitaev}.

This work is organized as follows. In Sec. I, we derive a fundamental lower bound on the number of physical fermion qubits $N$ required to encode $k$ logical qubits, such that the smallest logical operator on an encoded qubit is a $d$-body operator.  We refer to such fermion code as $[[N, k, d]]_{\boldsymbol{f}}$ code.  In Sec. II, we introduce the $[[6,1,3]]_{\boldsymbol{f}}$ code, which is the \emph{shortest} fermion quantum error-correcting code that encodes one logical qubit.  This code corrects quasiparticle poisoning errors as well as certain parity-conserving errors, and is thus applicable to systems of Majorana fermions and of ordinary fermions. In Sec. III, we demonstrate that any fermion code may be constructed from a restricted class of \emph{classical} binary error-correcting codes.  This correspondence allows us to introduce large families of non-degenerate  fermion codes, which we discuss at length.   We introduce a family of translationally-invariant codes, including codes with only four-body stabilizer operators. We present another family of codes in which logical qubit states have the same global fermion parity, thus permitting a simpler implementation of logical operations.

In Sec. IV, we reveal the significant advantages of error-correction in Majorana fermion systems, as compared to systems of bosons or ordinary fermions. Provided that Majorana fermions are well separated, the dominant error source is quasiparticle poisoning.  Interestingly, we find that in a sufficiently large system, error correction is possible when the poisoning probability is below a threshold ($\sim 10\%$) or {\it above} an upper limit ($\sim 90\%$). The remarkable feasibility of error correction in a very noisy environment stems from the fact that poisoning the same Majorana fermion twice recovers the qubit, so that error-correction in the Majorana platform is quasi-classical.
In Sec. V, we conclude by discussing the physical implementation of Majorana fermion error-correcting codes with four-body operators.

\section{Fermion Error-Correcting Codes}
We would like to identify efficient error-correcting codes for $N$ microscopic fermions that stabilize qubits encoded in the fermion parity.  Any complex fermion creation/annihilation operator can be expressed in terms of real (Majorana) fermion operators $\{\gamma_{n}\}$, which satisfy a very simple anti-commutation relation
\begin{align}\label{eq:majorana}
\{\gamma_{n}, \gamma_{m}\} = 2\delta_{nm}.
\end{align}
In this work, we consider a wide class of fermion error-correcting codes, where stabilizers $\mathcal{O}_{n}$ are products of an even number of Majorana fermion operators, hereafter referred to as ``Majorana stabilizers". These stabilizers mutually commute ($[\mathcal{O}_{n},\mathcal{O}_{m}] = 0$), square to the identity ($\mathcal{O}_{n}^{2} = +1$), and commute with the total fermion parity $\Gamma \equiv i^{N}\gamma_{1}\gamma_{2}\cdots\gamma_{2N}$. 

The unintentional action of the environment on the system, such as quasiparticle poisoning, can lead to decoherence. These errors are represented by the action of $t$-body operators on the code state. For example, a local fermion-parity flip corresponds to acting a single fermion operator on the qubits, hence $t=1$. In general, a ``weight-$t$ error" is correctable if it is has a unique syndrome, i.e. if the error is \emph{uniquely} identifiable through measurements of stabilizers, i.e. the operator acting on the physical qubits in a correctable error process anti-commutes with a unique set of stabilizer operators. The resulting stabilizer eigenvalue flips are the ``syndrome''.  Such a code, where each correctable error has a unique syndrome, is referred to as ``non-degenerate", and we will restrict our attention to such codes for important reasons that we elaborate at the end of this section.  Any fermion code in the remainder of our discussion is assumed to be non-degenerate, unless otherwise specified.

We refer to any fermion error-correcting code as an $[[N,k,d]]_{\boldsymbol{f}}$ code if $k$ qubits are encoded in $N$ complex fermions or $2N$ Majorana fermions, such that a $d$-body Majorana operator is the smallest logical operator acting on the encoded states, i.e. $d$ is the ``code distance".
If $t$ is the maximum weight of a correctable error in a non-degenerate $[[N, k, d]]_{\boldsymbol{f}}$ code, then the code distance $d \ge 2t + 1$.  A weight-$1$ error corresponds to an elementary quasiparticle poisoning event, as represented by a Majorana operator $\gamma_{j}$, which anti-commutes with the parity $\{\gamma_{j}, \Gamma\} = 0$.  In this language, the Kitaev Majorana chain (with stabilizers $\mathcal{O}_{n} = i\gamma_{2n}\gamma_{2n+1}$) is an $[[N, 1, 1]]_{\boldsymbol{f}}$ code, while four Majorana fermions with a single stabilizer ($\mathcal{O} \equiv \gamma_{1}\gamma_{2}\gamma_{3}\gamma_{4}$) define a $[[2, 1, 2]]_{\boldsymbol{f}}$ fermion code. Importantly, {neither} code can recover from quasiparticle poisoning events.  


We begin by formulating general conditions for fermion error-correcting codes with Majorana stabilizers. First, an $[[N, k, d]]_{\boldsymbol{f}}$ code must have exactly $N - k$ independent stabilizers to guarantee that the space of states satisfying $\mathcal{O}_{n}\ket{\Psi} = \ket{\Psi}$ is $2^k$-fold degenerate, which is used to encode $k$ logical qubits. Second, error correction on the encoded qubits requires constructing a mapping between operators whose action on the system generates errors and the $2^{N-k}-1$ stabilizer configurations describing states that result from the occurrence of errors.  For example, consider codes that are capable of correcting all elementary fermion-parity flip errors, as represented by single Majorana operators $\gamma_{1}, \ldots, \gamma_{2N}$. Clearly, it is possible to have a unique syndrome for each of these $2N$ error processes only if $2^{N-k}-1 \ge 2N$.  More generally, a non-degenerate $[[N, k, d]]_{\boldsymbol{f}}$ code exists, that can correct errors of weight less than $t$ only if
\begin{align}\label{eq:f_Hamming}
2^{N-k} \ge \sum_{m=0}^{t} \left(\begin{array}{c} 2N \\ m \end{array}\right).
\end{align}

Bounds similar to (\ref{eq:f_Hamming}) may also be derived for \emph{degenerate} fermion codes, for which two or more distinct, correctable errors have the same syndrome.  Such codes are of interest, as they would appear to be more efficient than non-degenerate codes.  We observe, however, that there are no fermion codes that are degenerate for weight-1 errors.  In such a code, at least two weight-1 operators (say $\gamma_{n}$ and $\gamma_{m}$) would have the same syndrome.  These errors would only be correctable if $i\gamma_{n}\gamma_{m} = +1$ in the codespace, making $\gamma_{n}$ and $\gamma_{m}$ ``ancilla" degrees of freedom that can then be removed entirely from the code. This argument demonstrates that any fermion code encoding one or more qubits must be non-degenerate for elementary quasiparticle poisoning (weight-1) errors.  Degenerate fermion codes that can correct for weight-2 errors, may also be of interest to correct for both de-phasing and quasiparticle poisoning in a system of complex fermions. Such codes only improve on the bound (\ref{eq:f_Hamming}) for {non-degenerate} fermion codes within a limited range of $N$, and the existence of these degenerate codes is not clear \cite{Forthcoming}.  Therefore, we choose to focus our attention on non-degenerate codes for the remainder of this work.

We conclude this section by observing that any Pauli stabilizer code may be trivially used to construct a fermion code, by representing each Pauli spin by four Majorana fermions with fixed fermion parity \cite{Kitaev_Spin_Liquid, Bravyi}.   
From any $[[N, k, d]]$ Pauli stabilizer code (encoding $k$ qubits in $N$ microscopic spins, with code distance $d$), one may construct a $[[2N, k, 2d]]_{\boldsymbol{f}}$ code which can correct weight-1 errors, by making the replacement $X_{n} \rightarrow i\chi^{(x)}_{n}\gamma_{n}$, $Y_{n} \rightarrow i\chi^{(y)}_{n}\gamma_{n}$,  and $Z_{n} \rightarrow i\chi^{(z)}_{n}\gamma_{n}$ for each Pauli spin, and by adding the stabilizer
$D_{n} \equiv i\chi^{x}_{n}\chi^{y}_{n}\chi^{z}_{n}\gamma_{n}$ at each site
to the code \cite{Kitaev_Spin_Liquid}.  Here, $\chi^{(x,y,z)}_{n}$ and $\gamma_{n}$ are Majorana fermion operators.  This mapping yields a class of fermion error-correcting codes that can correct for \emph{at most} weight-1 errors, and these codes are often not very efficient.  We shall soon present more efficient fermion codes, which cannot be mapped onto a bosonic code in this manner. 

\section{The Shortest Fermion Code}
In the following section, we introduce and study certain families of fermion error-correcting codes.  We begin by introducing the shortest fermion code $[[6,1,3]]_{\boldsymbol{f}}$, which encodes a single logical qubit and corrects for elementary fermion-parity flip errors. We thoroughly describe two operational modes for this error-correcting code in systems of either well-separated Majorana fermions or ordinary complex fermions.

Our $[[6,1,3]]_{\boldsymbol{f}}$ code is defined by the following stabilizers
\begin{align}
\mathcal{O}_{1} &= \gamma_{1}\gamma_{2}\gamma_{3}\gamma_{4} \nonumber \\
\mathcal{O}_{2} &= \gamma_{3}\gamma_{4}\gamma_{5}\gamma_{6} \nonumber \\
\mathcal{O}_{3} &= \gamma_{7}\gamma_{8}\gamma_{9}\gamma_{10} \nonumber \\
\mathcal{O}_{4} &= \gamma_{9}\gamma_{10}\gamma_{11}\gamma_{12}\nonumber \\
\mathcal{O}_{5} &= i\gamma_{2}\gamma_{4}\gamma_{6}\gamma_{8}\gamma_{10}\gamma_{12}
\end{align}
This code encodes a single fermion logical qubit, which can be represented by two logical Majorana fermion operators $\Gamma_{1,2}$. The fermion parity of this logical qubit, $i\Gamma_1 \Gamma_2$, is given by the total fermion parity of the system:
\begin{align}
i \Gamma_1 \Gamma_2 = \prod_{n=1}^{12}\gamma_{n},
\end{align}
and the logical parity flip operators $\Gamma_{1,2}$ are given by
\begin{align}
\Gamma_1 &= \gamma_{1}\gamma_{3}\gamma_{5}, \nonumber \\
\Gamma_2 &= \gamma_2 \gamma_4 \gamma_6 \gamma_7 \gamma_8 \gamma_9 \gamma_{10} \gamma_{11} \gamma_{12}.
\end{align}
These logical parity flip operators commute with all stabilizers.

Each elementary quasiparticle poisoning process anti-commutes with a unique combination of stabilizers, as indicated in the table below:
\begin{align*}
  \begin{tabular}{|c|c||c|c||}
  \hline
    $\,\gamma _{1}\,$ & $\mathcal{O}_{1}$ & $\,\gamma _{7}\,$ & $\mathcal{O}_{3}$ \\
    \hline
 $\gamma _{2}$ & $\mathcal{O}_{1},\mathcal{O}_{5}$ & $\gamma _{8}$ & $\mathcal{O}_{3},\mathcal{O}_{5}$ \\
     \hline
 $\gamma _{3}$ & $\mathcal{O}_{1},\mathcal{O}_{2}$ & $\gamma _{9}$ & $\mathcal{O}_{3}, \mathcal{O}_{4}$ \\
     \hline
 $\gamma _{4}$ & $\,\mathcal{O}_{1},\mathcal{O}_{2},\mathcal{O}_{5}\,$ & $\gamma _{10}$ & $\,\mathcal{O}_{3},\mathcal{O}_{4},\mathcal{O}_{5}\,$ \\
     \hline
      $\gamma _{5}$ & $\mathcal{O}_{2}$ & $\gamma _{11}$ & $\,\mathcal{O}_{4}\,$\\
     \hline
      $\gamma _{6}$ & $\,\mathcal{O}_{2},\mathcal{O}_{5}\,$ & $\gamma _{12}$ & $\,\mathcal{O}_{4},\mathcal{O}_{5}\,$ \\
     \hline
  \end{tabular}
\end{align*}
so that the code is non-degenerate for all weight-1 errors.  This may be used to decode and correct for poisoning events in the system that result bit-flip errors on the encoded qubit, by performing constant projective measurements of the commuting stabilizers.



In a system consisiting of well-separated Majorana fermions, dominant error processes are local quasiparticle poisoning events involving single Majorana fermions.  In contrast, for a system of  \emph{complex} fermions denoted by $c_j$, there can also be dephasing errors resulting from unintentional coupling of local fermion density $c_j^\dagger c_j$ to the environment.  Our $[[6,1,3]]_{\boldsymbol{f}}$ code also serves as an error-correcting code for \emph{complex} fermions, after pairing the twelve Majorana fermions into six complex fermions as follows
\begin{align}
c_{1} \equiv \frac{1}{2}(\gamma_{1} + i\gamma_{12})
\end{align}
and
\begin{align}
c_{n} \equiv \frac{1}{2}(\gamma_{n} + i\gamma_{n+5})\hspace{.2in}\text{for}\,\,n=2,\ldots,6
\end{align}

With the above identification, each of the local fermion parity operators $P_{n} \equiv 2c^{\dagger}_{n}c_{n} - 1$ anti-commutes with a unique combination of stablizers, as summarized in the following table:
\begin{align*}
  \begin{tabular}{|c|c||c|c||}
  \hline
    $\,P_{1}\,$ & $\mathcal{O}_{1}$,\,$\mathcal{O}_{4}$,\,$\mathcal{O}_{5}$ & $\,P_{2}\,$ & $\mathcal{O}_{1}$,\,$\mathcal{O}_{3}$,\,$\mathcal{O}_{5}$\\
    \hline
 $\,P_{3}\,$ & $\mathcal{O}_{1}$,\,$\mathcal{O}_{2}$,\,$\mathcal{O}_{3}$,\,$\mathcal{O}_{5}$ & $\,P_{4}\,$ & $\mathcal{O}_{1}$,\,$\mathcal{O}_{3}$,\,$\mathcal{O}_{4}$,\,$\mathcal{O}_{5}$\\
    \hline
   $\,P_{5}\,$ & $\mathcal{O}_{2}$,\,$\mathcal{O}_{3}$,\,$\mathcal{O}_{4}$,\,$\mathcal{O}_{5}$ & $\,P_{6}\,$ & $\mathcal{O}_{2}$,\,$\mathcal{O}_{4}$,\,$\mathcal{O}_{5}$\\
    \hline
   \end{tabular}
\end{align*}
Consider a de-phasing error, which takes the general form of a unitary operator $U_{n}(\delta\tau) = e^{i\delta\tau\,\sum_{n}t_{n}c^{\dagger}_{n}c_{n}}$ acting on the system.  For sufficiently short measurement times $\delta\tau$, we expand the unitary to linear order
$U_{n}(\delta\tau) = 1 + i\delta\tau\sum_{n}t_{n}c^{\dagger}_{n}c_{n} + O(\delta\tau^{2})$ and only consider \emph{on-site} de-phasing.
After the action of $U_{n}(\delta\tau)$ on the system, a measurement of the stabilizers will either project onto the original state of the system, or onto the state $P_{n}\ket{\Psi}$ up to an overall phase.  Since each $P_{n}$ has a unique syndrome, the de-phasing error may be uniquely determined and de-coded.  Bit-flip errors generated by elementary quasi-particle poisoning processes may also be decoded in this setup, as shown in the previous section. 

\section{General Framework}
We now introduce a general framework that allows us to systematically construct a large range of fermion codes, by revealing a remarkable connection between fermion codes with Majorana stabilizers and certain classical error-correcting codes, which allows for a systematic and efficient way to search for fermion codes.  This provides one of the main results of this work, and is presented in Sec. IIIA.   Based on this framework, we obtain two representative families of fermion error-correcting codes, which have larger code distance and can correct for higher-weight errors.   

\subsection{Constructing Fermion Codes from Classical Error Correcting Codes}
We now present a systematic construction of fermion codes, by relating them to certain \emph{classical} error-correcting codes, and reducing the search for fermion codes to a well-defined mathematical problem. The starting point for our construction is a representation of Majorana stabilizers as binary vectors.  Any product of Majorana fermion operators may be represented, up to an overall phase factor, as
\begin{align}
\mathcal{O}_{i} \sim \prod_{n=1}^{2N}(\gamma_{n})^{v^{(i)}_{n}}
\end{align}
where $\boldsymbol{v}^{(i)} \equiv ({v}^{(i)}_{1}, {v}^{(i)}_{2}, \ldots, {v}^{(i)}_{2N})$ is a \emph{binary} vector, i.e. a vector over the finite field $\mathbb{F}_{2}=\{0,1\}$, which is equipped with both $Z_{2}$ addition and $Z_2$ multiplication.  The product of operators $\mathcal{O}_{i}\mathcal{O}_{j}$ is represented as vector addition $\boldsymbol{v}^{(i)} + \boldsymbol{v}^{(j)}$.  An operator $\mathcal{O}_{j}$ commutes with the total fermion parity if and only if the corresponding binary vector is self-orthogonal
\begin{align}
\boldsymbol{v}^{(j)}\cdot(\boldsymbol{v}^{(j)})^{T} = 0,
\end{align}
and any pair of such operators mutually commute $[\mathcal{O}_{i}, \mathcal{O}_{j}] = 0$ if and only if
\begin{align}
\boldsymbol{v}^{(i)}\cdot(\boldsymbol{v}^{(j)})^{T} = 0.
\end{align}
In this notation, we may compactly specify any fermion error-correcting code by a binary ``stabilizer matrix"
\begin{align}
S \equiv \left( \begin{array}{c}
\boldsymbol{v}^{(1)}\\
\boldsymbol{v}^{(2)}\\
\vdots\\
\boldsymbol{v}^{(m)}
\end{array}\right)
\end{align}
satisfying
\begin{align}\label{eq:O_group}
S \cdot S^{T} = 0.
\end{align}
 While (\ref{eq:O_group}) is a necessary and sufficient condition for having independent, commuting stabilizers, the resulting fermion code may be unable to correct for any errors.  The distance of the fermion code is precisely the weight of the lowest-weight, binary vector $\boldsymbol{v}$ that satisfies $S\cdot \boldsymbol{v}^{T} = 0$, as $\boldsymbol{v}$ then corresponds to the smallest operator that commutes with all of the stabilizers and acts non-trivially on the codespace. 

A binary matrix satisfying (\ref{eq:O_group}) specifies a so-called ``weakly self-dual" binary \emph{classical} error-correcting code \cite{Classical_Coding}.  In a classical code, a bit-string, as specified by a binary vector $\boldsymbol{w}= (w_{1}, w_{2}, \ldots)$, may be encoded by multiplying by the \emph{generator matrix}
\begin{align}
G = \left( \begin{array}{c}
\boldsymbol{v}^{(1)}\\
\vdots\\
\boldsymbol{v}^{(m)}
\end{array}\right)
\end{align}
 of the classical code, i.e. $\boldsymbol{w}^{T} \rightarrow \boldsymbol{w}^{T} \cdot G$.  The space of valid encoded bit-strings is referred to as the \emph{codespace} $\mathcal{C} \equiv \mathrm{span}(\boldsymbol{v}^{(1)},\ldots,\boldsymbol{v}^{(m)})$, and any bit-string in the codespace is referred to as a codeword. A \emph{parity-check matrix} $H$, which projects onto the space orthogonal to $\mathcal{C}$, may be used to decode errors of sufficiently small weight on the encoded bit-string, since $H\cdot \boldsymbol{v} = 0$ if and only if $\boldsymbol{v}$ is a valid codeword (i.e. $\boldsymbol{v}\in \mathcal{C}$).  We may also construct the \emph{dual} code $\mathcal{C}^{\perp}$, with generator matrix $H$ and parity-check matrix $G$.   The condition $G\cdot G^{T} = 0$ implies that $\mathcal{C} \subseteq\mathcal{C}^{\perp}$, which defines a ``weakly self-dual" classical code.

\begin{table*}\label{table:Polynom}
  \begin{tabular}{|c||c||c|}
  \hline
    $\,\,\mathbf{N}\,\,$ & $\mathbf{f(x)}$ & \,\,{\bf Fermion Code} \,\,\\
    \hline
    $\,7\,$ & $1 + x + x^2 + x^4$ & $[[7,1,3]]_{\boldsymbol{f}}$\\
    \hline
    $\,14\,$ & $1 + x + x^4 + x^5 + x^6 + x^7$ & $[[7,0,4]]_{\boldsymbol{f}}$\\
    \hline
    $\,15\,$ & $1 + x + x^2 + x^3 + x^5 + x^7 + x^8 + x^{11}$ & $[[15,7,3]]_{\boldsymbol{f}}$\\
    \hline
    $\,21\,$ &
    $\begin{array}{c}
    1 + x^6 + x^{9} + x^{12} \\
    1 + x + x^{3} + x^{6} + x^{7} + x^{10} + x^{13} + x^{15}\\
    1 + x + x^{3} + x^{5} + x^{9} + x^{10} + x^{11} + x^{12}
    \end{array}$ &
    $\begin{array}{c}
    {[[}21,3,3{]]}_{\boldsymbol{f}} \\
    {[[}21,9,3{]]}_{\boldsymbol{f}} \\
    {[[}21,3,5{]]}_{\boldsymbol{f}}
    \end{array}$\\
    \hline
    $\,23\,$ & $1 + x + x^2 + x^3 + x^4 + x^7 + x^{10} + x^{12}$ & ${[[}23,1,7{]]}_{\boldsymbol{f}}$\\
    \hline
    $\,28\,$ &
    $\begin{array}{c}
     1 + x^4 + x^{8} + x^{16} \\
    1 + x + x^{3} + x^{4} + x^{5} + x^{7} + x^{8} + x^{9} + x^{11} + x^{16} + x^{19}\\
    1 + x^{2} + x^{8} + x^{10} + x^{12} + x^{14}\\
     1 + x^{2} + x^{4} + x^{7} + x^{8} + x^{9} + x^{11} + x^{15}\\
     1 + x + x^{2} + x^{3} + x^{4} + x^{5} + x^{7} + x^{10} + x^{11} + x^{12} + x^{15} + x^{16}\\
     1 + x + x^{2} + x^{5} + x^{8} + x^{9} + x^{10} + x^{12} + x^{14} + x^{17}\\
     1 + x^{3} + x^{5} + x^{6} + x^{8} + x^{11} + x^{12} + x^{13} + x^{14} + x^{15} + x^{17} + x^{18}\\
     1 + x + x^{3} + x^{4} + x^{5} + x^{7} + x^{8} + x^{9} + x^{11} + x^{16} + x^{17} + x^{19}
      \end{array}$ &
    $\begin{array}{c}
    {[[}14,2,3{]]}_{\boldsymbol{f}} \\
    {[[}14,5,3{]]}_{\boldsymbol{f}} \\
    {[[}14,0,4{]]}_{\boldsymbol{f}} \\
    {[[}14,1,4{]]}_{\boldsymbol{f}} \\
    {[[}14,2,4{]]}_{\boldsymbol{f}} \\
    {[[}14,3,4{]]}_{\boldsymbol{f}} \\
    {[[}14,4,4{]]}_{\boldsymbol{f}} \\
    {[[}14,5,3{]]}_{\boldsymbol{f}}
    \end{array}$\\
    \hline
    $\,30\,$ &
    $\begin{array}{c}
     1 + x + x^{2} + x^{5} + x^{9} + x^{10} + x^{11} + x^{12} + x^{14} + x^{15}\\
    1 + x^{2} + x^{3} + x^{4} + x^{5} + x^{7} + x^{8} + x^{11} + x^{16} + x^{19}\\
    1 + x^{2} + x^{4} + x^{5} + x^{6} + x^{7} + x^{9} + x^{11} + x^{12} + x^{17}\\
     1 + x + x^{4} + x^{5} + x^{10} + x^{11} + x^{12} + x^{13} + x^{16} + x^{17} + x^{18} + x^{19}
     + x^{20} + x^{21}\\
     1 + x^{3} + x^{5} + x^{6} + x^{9} + x^{13} + x^{14} + x^{16}\\
     1 + x + x^{2} + x^{6} + x^{7} + x^{9} + x^{11} + x^{12} + x^{16} + x^{17} + x^{19} + x^{20}\\
     1 + x + x^{2} + x^{3} + x^{4} + x^{8} + x^{9} + x^{10} + x^{11} + x^{13} + x^{17} + x^{18}\\
     1 + x^{2} + x^{4} + x^{6} + x^{10} + x^{14} + x^{16} + x^{22}
      \end{array}$ &
    $\begin{array}{c}
    {[[}15,0,6{]]}_{\boldsymbol{f}} \\
    {[[}15,4,4{]]}_{\boldsymbol{f}} \\
    {[[}15,2,6{]]}_{\boldsymbol{f}} \\
    {[[}15,6,4{]]}_{\boldsymbol{f}} \\
    {[[}15,1,6{]]}_{\boldsymbol{f}} \\
    {[[}15,5,3{]]}_{\boldsymbol{f}} \\
    {[[}15,3,5{]]}_{\boldsymbol{f}} \\
    {[[}15,7,3{]]}_{\boldsymbol{f}}
    \end{array}$\\
    \hline
   \end{tabular}
   \caption{{\bf Translationally-Invariant Fermion Codes:} A list of the weakly self-dual (binary) classical codes of size $N \le 30$ with distance $d \ge 3$, and the corresponding fermion codes that they give rise to, using the mapping presented in the main text.  When $N$ is odd, we may obtain a fermion code in a variety of ways, as discussed in Sec. IIID.  The fermion codes presented in the table for odd $N$ are obtained from two copies of the classical code, so that the resulting  system has an even number of Majorana fermions, and describes a physical Hilbert space.}
\end{table*}

The relation we have derived allows us to view the generator matrix $G$ of any weakly self-dual classical code as the stabilizer matrix $S$ of a fermion code.  Let the weakly self-dual code have code parameters $[2N, k, d]$, where the classical code distance $d$ is precisely the minimum weight of a codeword, while the classical code dimension $k = \mathrm{dim}(\mathcal{C})$.  Our construction yields the following mapping to a fermion code
\begin{align}
[2N, k, d] \,\,\,\longrightarrow\,\,\,[[N,N-k,d^{\perp}]]_{\boldsymbol{f}}
\end{align}
where $d^{\perp}$ is the code distance of the dual code $\mathcal{C}^{\perp}$.  If the weakly self-dual code $\mathcal{C}$ involves an odd number of bits, then we can employ various schemes to construct a fermion code, which we elaborate on in Sec. IIID.

\subsection{Cyclic Fermion Codes}

We now construct representative families of fermion codes with code distance $d \ge 3$ by taking advantage of our mapping to weakly self-dual classical codes.  The first family includes all fermion codes with one-dimensional translational symmetry.  These codes are obtained from a subset of ``cyclic" codes in classical coding theory \cite{Classical_Coding}.  The second family describes an infinite set of codes with increasing distance, where all of the encoded qubits have the same global fermion parity.  This property would theoretically permit a simpler practical implementation of the logical operators in the code.

We may construct one-dimensional, translationally-invariant fermion codes through the following procedure, which we prove in the supplemental material \cite{SM}.   Consider the polynomial ring $\mathbb{F}_{2}[x]$, i.e. the set of polynomials in $x$ with coefficients in the field $\mathbb{F}_{2}$.  We now
\begin{enumerate}[label=\Roman*.]
\item Factorize $x^{N}-1$ as $x^{N}-1 = f(x)g(x)$\\
\item Check that $\widetilde{f}(x) \equiv x^{\mathrm{deg}(g)}g(x^{-1})$ divides $f(x)$,\\
\hspace{.2in}where, $\mathrm{deg}(g)$ is the degree of $g(x)$.
\end{enumerate}
The first condition provides a well-known way to generate translationally-invariant (cyclic) classical error-correcting codes \cite{Classical_Coding}, while the additional second condition yields the subset of these codes which are weakly self-dual, as we demonstrate in the supplemental material \cite{SM}.  If a polynomial $f(x)$ satisfies both conditions, then it may be used to define a translationally-invariant fermion code in the following manner. From the polynomial $f(x)$, which we write as
\begin{align}\label{eq:polynom}
f(x) = \sum_{m=0}^{N-1}f_{m}x^{m}
\end{align}
we extract the vector $\boldsymbol{f} = (f_{0},\ldots, f_{N-1})$, which we may take to be the binary vector representation of a Majorana stabilizer.  In this way, any polynomial of the form (\ref{eq:polynom}) represents a Majorana operator acting on a system of $N$ Majorana fermions.  Similar techniques have been successfully applied to find Majorana stabilizer codes in two and three spatial dimensions \cite{Vijay_Haah_Fu_1}.

If $N$ is even, we take the polynomials
\begin{align}
f(x),\,\,x\,f(x),\,\,x^{2}f(x),\,\ldots,\,\,x^{N - \mathrm{deg}(f)}f(x)
\end{align}
to define the $[N - \mathrm{deg}(f)]$ stabilizers of a fermion code with $N$ microscopic Majorana fermions.  These stabilizers are all independent of each other, by construction, while the condition (II) guarantees that all of the stabilizers commute with each other and with the fermion parity, as shown in the Supplemental Material \cite{SM}.  In practice, $f(x)$ is precisely the so-called ``generator polynomial" \cite{Classical_Coding} of a weakly self-dual, binary cyclic classical code, which we have then used to define a fermion code.
If $N$ is odd, we may take two disjoint copies of the classical code to define the resulting fermion code, which now has $2N$ Majorana fermions.  In this case, the fermion code is defined by the stabilizers
\begin{align}
& f(x),\,\,x\,f(x),\,\,x^{2}f(x),\,\ldots,\,\,x^{N - \mathrm{deg}(f)}f(x),\\
& x^{N}f(x),\,\,x^{N+1}\,f(x),\,\ldots,\,\,x^{2N - \mathrm{deg}(f)}f(x)\nonumber
\end{align}

Translationally-invariant fermion codes with code distance $d \ge 3$ that are obtained from weakly self-dual cyclic codes of size $N \le 30$ via our construction, are indicated in Table I.  Many of these codes can correct for higher-weight errors, beyond the simplest quasiparticle poisoning events.  For example, the $[[15,1,6]]_{\boldsymbol{f}}$ code has code-distance $d = 6$ and can correct for any weight-$2$ fermionic error.  From the polynomial $f(x)$ indicated in Table I, we observe that the fermion code is defined by stabilizers $\mathcal{O}_{1}, \mathcal{O}_{2},\ldots, \mathcal{O}_{14}$ where
\begin{align}
\mathcal{O}_{i} \equiv \gamma_{i}\gamma_{i+3}\gamma_{i+5}\gamma_{i+6}\gamma_{i+9}\gamma_{i+13}\gamma_{i+14}\gamma_{i+16}
\end{align}
This code may be used either for well-separated Majorana fermions or for complex fermions, to correct elementary de-phasing or quasiparticle poisoning errors.  As another example, the
$[[23,1,7]]_{\boldsymbol{f}}$ code, which is based on the well-known classical Golay code, has stabilizers $\mathcal{O}_{1},\mathcal{O}_{2},\ldots,\mathcal{O}_{11}$, $\mathcal{O}_{23},\mathcal{O}_{24},\ldots, \mathcal{O}_{33}$ where
\begin{align}
\mathcal{O}_{i} \equiv \gamma_{i}\gamma_{i+1}\gamma_{i+2}\gamma_{i+3}\gamma_{i+4}\gamma_{i+7}\gamma_{i+10}\gamma_{i+12}
\end{align}
and can correct for any weight-$3$ fermionic error.


\subsection{Fermion Codes with Fixed Global Fermion Parity in the Codespace}
More complex families of fermion error-correcting codes with increasing code distance may also be constructed by searching for other weakly self-dual classical codes. We now review the code parameters for one such family of fermion codes, which has the important advantage that the global fermion parity is fixed in the codespace, so that all of the encoded qubits have the same global parity.  In practice, this would make practical implementation of these codes more feasible, since all logical operators on the encoded qubits simply measure the parity of some subset of the Majorana fermions in the code.  The explicit construction of the stabilizers for these codes is provided in the Supplemental Material \cite{SM}.  This family of fermion codes has code parameters
\begin{align}\label{eq:RM_Code_f}
\left[\left[2^{m-1}, \,\,\,2^{m-1} - B(r,m),\,\,\,\,2^{r+1}\right]\right]_{\boldsymbol{f}}
\end{align}
with $m \ge 2r+1$ and
\begin{align}
B(r,m) \equiv \displaystyle\sum_{j=0}^{r}\left(\begin{array}{c} m\\ j \end{array}\right).
\end{align}
We refer to this family as the RM$_{\boldsymbol{f}}(r,m)$ codes as these codes are obtained from a subset of the well-known Reed-Muller classical error-correcting codes.  Any member of this family can \emph{at least} correct for elementary quasiparticle poisoning errors.

 \begin{figure}
 \includegraphics[trim = 0 0 0 0, clip = true, width=0.5\textwidth, angle = 0.]{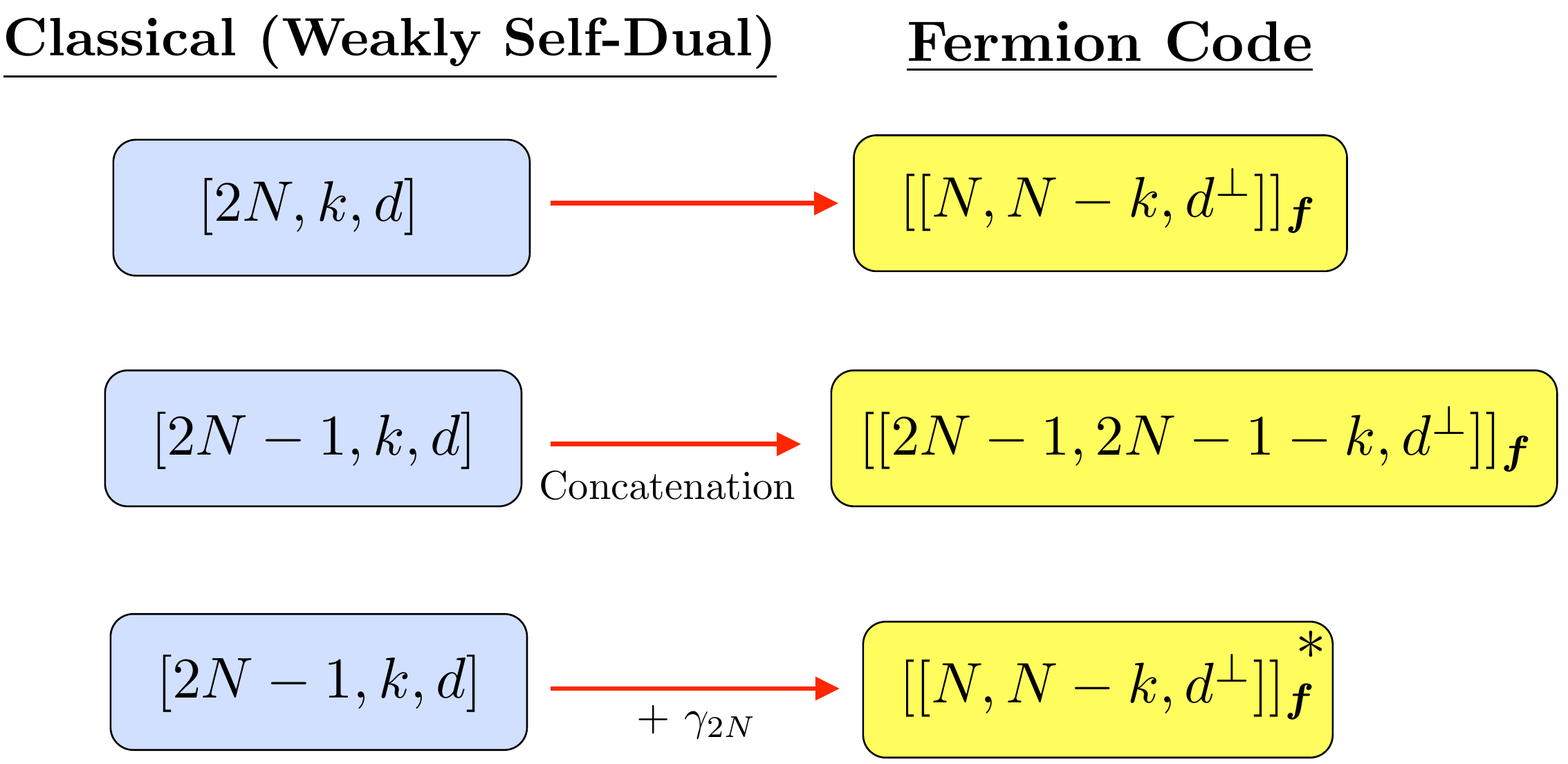}
 \caption{{\bf Code Mappings:} A summary of the mappings between classical (weakly self-dual) error-correcting codes, and various fermion codes.  When the number of bits in the classical code is even, the generator matrix of the code may be taken to be the stabilizer matrix $S$ of the fermion code.  When the number of bits is odd, one can concatenate two such codes, or add a single Majorana so as to describe a physical Hilbert space.  The latter case is useful in platforms with complex fermions where quasiparticle poisoning is suppressed, and the indicated code distance assumes that parity-violating processes are forbidden.}
  \label{fig:Code_Map}
\end{figure}

A simple example of a code in this family is RM$_{\boldsymbol{f}}(1,4) = [[8,3,4]]_{\boldsymbol{f}}$, whose stabilizers may be written as eight-Majorana operators, in the form
\begin{align}
&\mathcal{O}_{1} \equiv \prod_{m=1}^{8}\gamma_{2m} \hspace{.34in} \mathcal{O}_{2} \equiv \prod_{m=1}^{8}\gamma_{2m-1}\\
&\mathcal{O}_{3} \equiv \prod_{m=1}^{8}\gamma_{m} \hspace{.34in} \mathcal{O}_{4} \equiv \prod_{m=1}^{4}\gamma_{m}\gamma_{m+9}\\
&\hspace{.34in}\mathcal{O}_{5} \equiv \prod_{m=1}^{2}\gamma_{m}\gamma_{m+4}\gamma_{m+8}\gamma_{m+12}
\end{align}
Observe that the global fermion parity $\Gamma = \mathcal{O}_{1}\mathcal{O}_{2}$, so that all of the encoded states have parity $\Gamma = +1$.  \\

\subsection{Fermion Codes and Code Concatenation}
We now present, in formal terms, the more general schemes for constructing fermion codes from weakly self-dual classical error correcting codes with an \emph{odd} number of bits, i.e. with code parameters $[2N + 1, k, d]$ and generator matrix $G$.  In this case, the simplest way to construct a fermion code in order to guarantee that the resulting code has an even number of Majorana zero modes, and thus describes a physical Hilbert space, is to view $G \oplus G$ as the stabilizer matrix for a fermion code.  This ``code concatenation" protocol yields the mapping
\begin{align}\label{eq:Map_2}
[2N-1, k, d] \,\,\,\longrightarrow\,\,\,[[2N-1,2N-1-k,d^{\perp}]]_{\boldsymbol{f}}
\end{align}
and is reminiscent of the well-known Calderbank-Shor-Steane (CSS) construction  \cite{CSS, Shor_Code}, in which two classical error correcting codes $\mathcal{C}_{1,2}$ satisfying $\mathcal{C}_{1}\subseteq\mathcal{C}_{2}$, may be used to construct a Pauli stabilizer code.  The mapping (\ref{eq:Map_2}) is equivalent to the statement that a weakly self-dual CSS code may be used to construct a fermion code by replacing $X_{i}$, $Z_{i} \rightarrow\gamma_{i}$ \cite{Bravyi}.   This construction may also be generalized by ``concatenating" two different weakly self-dual classical codes, each with an odd number of bits, to produce a new fermion code.

A second protocol -- which is useful as a method for constructing codes in platforms with complex fermions in which de-phasing is the primary source of error -- involves taking $G$ to be the stabilizer matrix for a code in a platform of $2N$ Majorana fermions.  Since the classical code  involves $2N-1$ bits, none of the stabilizers in the fermion code act on the last Majorana fermion ($\gamma_{2N}$).  As a result, while not all elementary quasiparticle poisoning errors are detectable, higher-weight errors can still be corrected, a feature which may be useful in platforms where de-phasing is the primary error source and quasiparticle poisoning is suppressed. We emphasize that both of the above protocols are only needed in the construction of a fermion code when the classical code involves an odd number of bits.

\section{Comparing Error Correction in Bosonic and Fermionic Codes}
We conclude our discussion of fermion codes by highlighting the important differences between error correction in Bose and Fermi systems, as well as the advantages of error correction in fermionic platforms where quasiparticle poisoning is the dominant error-source.  As we have emphasized, there Fermi statistics generically forbids a local unitary mapping between fermionic and bosonic systems.  In contrast, a non-local unitary transformation on a Bose system will not give rise to a fermion error-correcting code, since local errors in the fermion system will be non-local in the Bose system and will be un-correctable.  An example of this phenomenon is given in the supplemental material \cite{SM}.

Our approach to fermion codes highlights an important advantage of error correction in Majorana platforms, where quasiparticle poisoning is the dominant error source.  Our inequality for non-degenerate Majorana stabilizer codes (\ref{eq:f_Hamming}) is the fermionic counterpart of the quantum Hamming bound for Pauli stabilizer codes \cite{Hamming_Saturate, Gottesman_Thesis}.  The asymptotic limit of the Hamming bound for large $N$ \cite{CSS} highlights an important advantage of error-correction in the Majorana platform.
Let $p$ be the probability per site that an elementary quasiparticle poisoning error occurs within a time $\Delta\tau$.  The probability $p$ gives the \emph{typical} fraction of Majorana fermions that have experienced a quasiparticle poisoning event in this time interval; in other words, an error of weight $2Np$ will have typically occurred in time $\Delta \tau$.  Assuming that our code only needs to correct for typical error configurations, the inequality (\ref{eq:f_Hamming}) simplifies, and its behavior for large $N$ is given by
\begin{align}
\epsilon \le 1 - 2\,H(p)
\end{align}
where $\epsilon \equiv k/N$ is the \emph{code efficiency} and $H(p)$ is the binary entropy function $H(p) \equiv -p\log_{2}(p) - (1-p)\log_{2}(1-p)$.  Error-correction appears to possible in a fermion code  even when the poisoning probability $p$ is \emph{large}, as the code efficiency $\epsilon > 0$ when $|p - \frac{1}{2}| > c$ with $c \approx 0.39$. This reflects the fact that the number of typical configurations for quasiparticle poisoning in a fermion system decrease as the error probability $p \rightarrow 1$. In contrast, the entropy of error configurations grows much faster in Pauli stabilizer codes, since there are three single-qubit errors per site, giving rise to a probability threshold $p_{c}$ above which error correction is impossible.

\section{Physical Implementation of Fermion Codes}
Fermion codes with sufficiently few-body interactions may admit a convenient physical implementation in platforms with Majorana zero modes.   As demonstrated in \cite{Teleportation} a striking signature of Majorana fermions in a mesoscopic superconducting island is the presence of {phase-coherent} single-electron transport (termed ``teleportation").  Measurements of the transmission phase-shift of electron ``teleportation" through Majorana zero modes -- by measuring the conductance in an electron interferometer or the persistent current in
a closed loop -- may be used to perform projective measurements of two-body and four-body Majorana operators \cite{Braiding_without_Braiding}.

In this section, we present a physical implementation of the simplest fermion code derived from weakly self-dual Reed-Muller codes in Sec. IIC, RM$_{\boldsymbol{f}}(1,3) = [[4,0,4]]_{\boldsymbol{f}}$ code, which is defined by the stabilizers
\begin{align}
\mathcal{O}_{1} \equiv \gamma_{1}\gamma_{3}\gamma_{5}\gamma_{7} \hspace{.3in}
\mathcal{O}_{2} \equiv \gamma_{2}\gamma_{4}\gamma_{6}\gamma_{8}\\
\mathcal{O}_{3} \equiv \gamma_{3}\gamma_{4}\gamma_{5}\gamma_{6} \hspace{.3in} \mathcal{O}_{4} \equiv \gamma_{5}\gamma_{6}\gamma_{7}\gamma_{8}
\end{align}
While the codespace encodes no qubits, this code can correct for any elementary quasiparticle poisoning errors; as a result, its physical implementation may be useful to study quasiparticle poisoning times in a platform of Majorana zero modes.  Our physical implementation may be extended to implement other fermion codes with sufficiently few-body interactions, such as the translationally-invariant $[[7,1,3]]_{\boldsymbol{f}}$ code, which only involves quartic interactions, and is the fermionic counterpart to the well-known Steane code \cite{Steane_Code}. 

 \begin{figure}
 $\begin{array}{c}
 \includegraphics[trim = 0 0 0 0, clip = true, width=0.37\textwidth, angle = 0.]{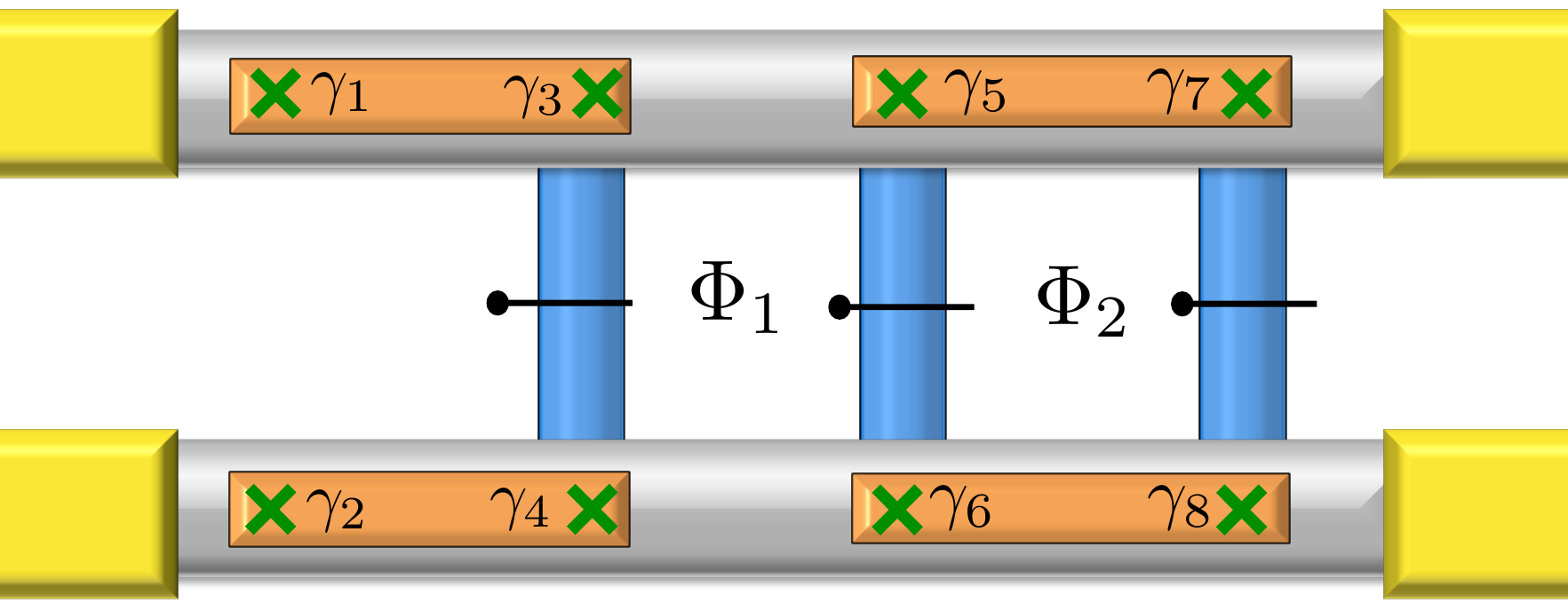}\\
 \text{(a) Measurement of $\mathcal{O}_{1}$ and $\mathcal{O}_{2}$}\\\\
  \includegraphics[trim = 0 0 0 0, clip = true, width=0.37\textwidth, angle = 0.]{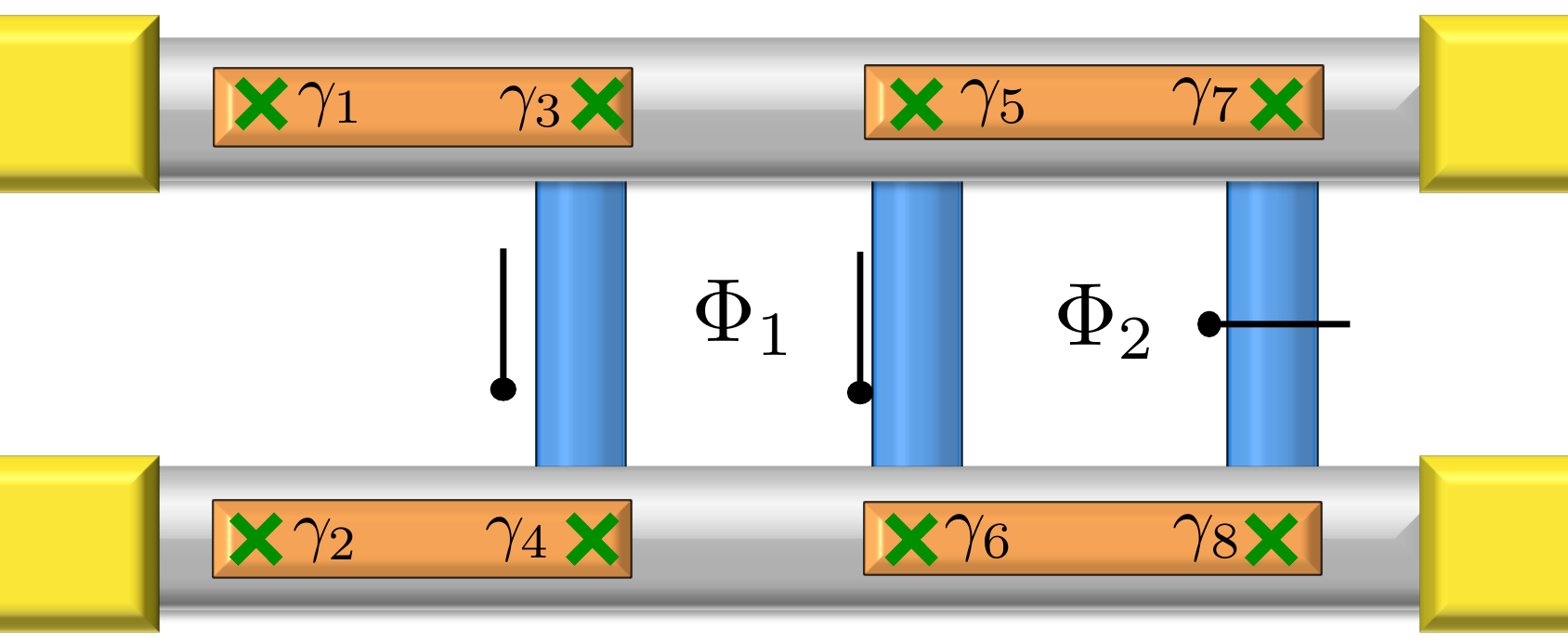}\\
  \text{(b) Measurement of $\mathcal{O}_{3}$}\\
 \end{array}$
 \caption{{\bf Implementation of the $[[4,0,4]]_{f}$ Code:} Gate voltages are applied on two proximitized nanowires so that 6 well-separated Majorana fermions appear.  Metal bridges (shown in blue) couple the nanowires, with voltages applied to tune the couplings ``on" or ``off" as indicated by the switches in the figures. Each wire has a large charging energy. In (a), the two-terminal conductance using the leads (shown in yellow) may be used to measure $\mathcal{O}_{1}$ and $\mathcal{O}_{2}$, while in (b) a measurement of the persistent current flowing in a loop enclosing flux $\Phi_{1}$ measures $\mathcal{O}_{3}$.  A similar protocol involving a flux $\Phi_{2}$ may be used to measure $\mathcal{O}_{4}$.  }
  \label{fig:Implementation}
\end{figure}

The physical implementation of the $[[4,0,4]]_{\boldsymbol{f}}$ code involves using two parallel semicoductor nanowires with spin-orbit coupling, and proximitized by $s$-wave superconductors, as shown in Fig. \ref{fig:Implementation}.  In the presence of a large Zeeman field, it is well-known that the proximitized nanowire goes into a topological superconducting regime, which localizes Majorana zero modes at the interface with a trivial superconductor.  In the setup shown in Fig. \ref{fig:Implementation}, gate voltages may be applied along the length of the nanowires to create alternating interfaces between topological and trivial $p$-wave superconducting segments that host Majorana zero modes, in order to realize the configuration of twelve Majorana fermions shown. The Majoranas are well-separated so that hybridization between adjacent zero modes may be neglected.  Furthermore, normal metal bridges (shown in blue) are placed between the two nanowires and between the lower nanowire to introduce a coupling between the appropriate Majorana zero modes, which may be tuned by applying voltages along the bridges.  Finally, we assume that the total charge on each nanowire segment is fixed by taking the charging energy $E_{C}$ to be large on each nanowire..

Due to the large charging energy on each nanowire, the quartic Majorana operators $\mathcal{O}_{1}$ and $\mathcal{O}_{2}$ in the $[[4,0,4]]_{\boldsymbol{f}}$ code may be measured by applying fluxes through elementary loops formed by the nanowires and the metal bridges in our setup and measuring the persistent current, which depends sensitively on the stabilizer eigenvalue (e.g. $I \sim \epsilon (e/\hbar)\sin[e\Phi_{1}/\hbar]\mathcal{O}_{1}$).  Projective measurements of the other two quartic operators may be performed by measuring the two-terminal conductance through leads attached to the ends of the two nanowires -- shown in yellow in Fig. \ref{fig:Implementation} -- which is sensitive to the eigenvalue of the appropriate quartic operator.  As an example, for the measurement of the $\mathcal{O}_{3}$ in Fig. \ref{fig:Implementation}b, the two-terminal conductance $G = g_{0} + g_{1}\mathcal{O}_{3}$ \cite{Braiding_without_Braiding, Egger}.  Intricate details of the persistent-current and conductance-based measurements of Majorana operators are provided in Ref. \cite{Braiding_without_Braiding}.

\section{acknowledgments}
This work is supported by DOE Office of Basic Energy Sciences, Division of Materials Sciences and Engineering under Award DE-SC0010526. LF is supported partly by the David and Lucile Packard Foundation. SV was supported for part of this work by the National Science Foundation under Grant No. NSF PHY11-25915.

\appendix


\section{Weakly Self-Dual Cyclic Codes}
The codespace $\mathcal{C}$ of a cyclic code on $N$ bits is identified with an \emph{ideal} $I$ in the quotient ring $R = \mathbb{F}_{2}[x]/\langle x^{N}-1\rangle$ \cite{Classical_Coding} where $\mathbb{F}_{2}[x]$ is the polynomial ring over the finite field $\mathbb{F}_{2}$, while
\begin{align}
\langle x^{N} - 1\rangle \equiv \left\{ h(x)\,(x^{N}-1)\,|\,h(x)\in\mathbb{F}_{2}[x]\right\}
\end{align}
is the ideal generated by $x^{N}-1$.  An ideal $I$ is a collection of elements $\{g_{i}\}$ in the ring $R$ satisfying the property that $\sum_{i}r_{i}g_{i}\in I$ where $r_{i}\in R$; that is, $I$ is an additive subgroup of $R$ that is invariant under multiplication by any element of $R$.

To construct the generator matrix for any weakly self-dual cyclic code, we first observe that $R$ is a principal ideal ring, i.e. any ideal $I \subseteq R$ is generated by a single polynomial $f(x)\in I$, which is the lowest-degree, monic polynomial contained in the ideal.  We may prove this by contradiction.  Let $h(x)\in I$, so that $h(x) = q(x)f(x) + r(x)$ where the remainder $r(x)\ne 0$ iff $\mathrm{deg}(r) < \mathrm{deg}(g)$.  Since $h(x)\in I$, we must have $r(x)\in I$, in which case $\mathrm{deg}(r) \ge \mathrm{deg}(g)$ since $f(x)$ is the lowest-degree monic polynomial in $I$.  Therefore, $r(x) = 0$ and $f(x)$ divides any $h(x)\in I$.  We conclude that $I = \langle f(x)\rangle$.

Furthermore, the generator $f(x)$ of any ideal in $R$ must divide $x^{N}-1$. We may again prove this by contradiction.  Let $q(x)f(x) + r(x) = x^{N}-1$ for some appropriate choice of $q(x)$ and $r(x)$, with $r(x)\ne 0$ iff $\mathrm{deg}(r) < \mathrm{deg}(g)$. As a result, $r(x) = q(x)f(x)$ mod $x^{N}-1$, so that $r(x)\in I$.  Using the fact that $f(x)$ has minimal degree in $I$, we conclude that $\mathrm{deg}(r) > \mathrm{deg}(g)$, so that $r(x) = 0$.  As a result, $f(x)$ divides $x^{N}-1$.

Therefore, the generator of any ideal $I\subseteq R$ may be constructed by factorizing $x^{N}-1 = f(x)g(x)$.  Since $f(x)$ is a monic polynomial of degree-$d$, $g(x)$ must be also be monic, of degree $(N-d)$.  If we write
\begin{align}
f(x) = \sum_{m=0}^{d}f_{m}x^{m} \hspace{.2in} g(x) = \sum_{m=0}^{N-d}g_{m}x^{m}
\end{align}
then the condition $x^{N}-1 = f(x)g(x)$ is equivalent to the matrix equation $F\cdot G^{T} = 0$ where
\begin{align}
&F = \left(\begin{array}{cccccccc}
f_{0} & f_{1} & f_{2} & \cdots & f_{d} & 0 & 0 & \cdots \\
0 & f_{0} & f_{1} & \cdots & f_{d-1} & f_{d} & 0 & \cdots \\
\vdots &  &  \ddots & \ddots &  & \ddots & \ddots &\\
0 & 0 & \cdots & f_{0} & f_{1} & \cdots & f_{d-1} & f_{d}
\end{array}\right)\\ \nonumber\\
&G = \left(\begin{array}{cccccccc}
g_{N-d} & g_{N-d-1} & \cdots & \cdots & g_{0} & 0 & 0 & \cdots \\
0 & g_{N-d} & \cdots & \cdots & g_{1} & g_{0} & 0 & \cdots \\
\vdots &  &  \ddots &  &  & \ddots & &\\
0 & 0 & \cdots & g_{N-d} & \cdots & \cdots & g_{1} & g_{0}
\end{array}\right)\nonumber
\end{align}
We observe that, by construction both $G$ and $F$ have full row-rank.  Therefore, if we take $F$ to be the generator matrix of the classical cyclic code $\mathcal{C}$, then $G$ must be the parity-check matrix.  Alternatively, $G$ is the generator matrix of the dual code $\mathcal{C}^{\perp}$, with generator polynomial
\begin{align}\label{eq:dual}
\widetilde{f}(x) \equiv x^{N-d}g(x^{-1})
\end{align}
The code $\mathcal{C}$ is weakly self-dual iff $\mathcal{C}\subseteq\mathcal{C}^{\perp}$.  This can only be the case if $\langle f(x) \rangle \subseteq \langle \widetilde{f}(x) \rangle$ in which case, $\widetilde{f}(x)$ must divide $f(x)$.

To summarize, we may factorize $x^{N}-1 = f(x)g(x)$ and then check that $\widetilde{f}(x)$, as defined in (\ref{eq:dual}), divides $f(x)$.  In this case, the matrix $F$ may be taken to be the binary \emph{stabilizer matrix} for the fermion code.  Since $F^{T}F = 0$ by construction, all of the operators in the code commute with each other and with the total fermion parity.

\section{RM$_{\boldsymbol{f}}(r,m)$ from Reed Muller Codes}
As we have already seen, a convenient way to construct a binary matrix $S$ satisfying $S\cdot S^{T} = 0$, is to construct the rows of $S$ from orthogonal basis vectors for a binary vector space.
A particularly convenient choice, which yields a class of fermion codes where the total fermion parity is fixed in the codespace is given by classical Reed-Muller codes RM$(r, m)$ \cite{Classical_Coding}, which are constructed from the vector space of polynomials of degree-$r$ in $m$ binary variables. The $\mathbb{F}_{2}$ dimension of this space is precisely
\begin{align}
k_{\mathrm{RM}} = \sum_{j = 0}^{r}\left(\begin{array}{c} m\\ j \end{array}\right).
\end{align}
The Reed-Muller codes are known to be weakly self-dual \cite{Classical_Coding} when
\begin{align}
m \ge 2r+1.
\end{align}
The microscopic expressions for the stabilizers in the corresponding fermion error-correcting code are obtained by representing each monomial of degree less than $(r+1)$ as a binary vector.  We begin by representing the variable $x_{k}$ as a binary vector $\boldsymbol{v}^{(k)}\in(\mathbb{F}_{2})^{2^{m}}$ with an alternating sequence of $2^{m-k}$ ones and zeros, i.e.
\begin{align}
x_{k} \longrightarrow \boldsymbol{v}^{(k)} \equiv (\,\underbrace{1,\ldots, 1,}_{2^{m-k}\,\text{times}\,\,\,}\underbrace{0,\ldots, 0,}_{2^{m-k}\,\text{times}\,\,\,} \underbrace{1,\ldots, 1,}_{2^{m-k}\,\text{times}\,}\ldots)\nonumber
\end{align}
while $1$ is represented as
\begin{align}
1 \longrightarrow \boldsymbol{v}^{(0)} \equiv (\,\underbrace{1,\ldots,1}_{2^{m}\,\text{times}})
\end{align}
Monomials are represented by taking vector products, i.e.
\begin{align}
x_{j}x_{k}\,\,\, \longrightarrow\,\,\, \boldsymbol{v}^{(j,k)} \equiv  \boldsymbol{v}^{(j)}\star  \boldsymbol{v}^{(k)}
\end{align}
where
\begin{align}
\boldsymbol{v}^{(j)}\star  \boldsymbol{v}^{(k)} \equiv \left(v^{(j)}_{1}v^{(k)}_{1},\,v^{(j)}_{2}v^{(k)}_{2},\ldots, v^{(j)}_{2^{m}}v^{(k)}_{2^{m}} \right)
\end{align}
In this way, we may represent all of the monomials that form the basis elements of the vector space of polynomials of degree-$r$ in $m$ binary variables, as binary vectors.  These vectors are then used to construct the rows of the generator matrix of the Reed-Muller code \cite{Classical_Coding}, which may then be taken to be the stabilizer matrix of a fermion code when $m \ge 2r+1$.

As an example, the matrix of stabilizers for the $[[4,0,4]]_{\boldsymbol{f}}$ code, as obtained from this construction, is precisely the generator matrix for RM$(1,3)$ and is given by
\begin{align}
S_{[[4,0,4]]_{\boldsymbol{f}}} = \left(\begin{array}{cccccccc}
1 & 1 & 1 & 1 & 1 & 1 & 1 & 1\\
1 & 1 & 1 & 1 & 0 & 0 & 0 & 0\\
1 & 1 & 0 & 0 & 1 & 1 & 0 & 0\\
1 & 0 & 1 & 0 & 1 & 0 & 1 & 0
\end{array}\right)
\end{align}
Appropriately adding rows of this matrix yields a more convenient choice of stabilizers  for this code, involving only quartic Majorana operators. These stabilizers may be explicitly written as
\begin{align}
&\mathcal{O}_{1} \equiv \gamma_{1}\gamma_{3}\gamma_{5}\gamma_{7}\hspace{.3in}\mathcal{O}_{2} \equiv \gamma_{2}\gamma_{4}\gamma_{6}\gamma_{8}\\
&\mathcal{O}_{3} \equiv \gamma_{3}\gamma_{4}\gamma_{5}\gamma_{6}\hspace{.3in}\mathcal{O}_{4} \equiv \gamma_{5}\gamma_{6}\gamma_{7}\gamma_{8}.
\end{align}
as presented in the main text.  Similarly, the matrix of stabilizers for RM$_{\boldsymbol{f}}(1,4) = [[8,3,4]]_{\boldsymbol{f}}$ is given by
\begin{align}
S_{[[8,3,4]]_{\boldsymbol{f}}} = (\boldsymbol{v}^{(0)}, \boldsymbol{v}^{(1)},\boldsymbol{v}^{(2)}, \boldsymbol{v}^{(3)}, \boldsymbol{v}^{(4)})^{T}
\end{align}
Yet another example is given by RM$_{\boldsymbol{f}}(2,5) = [[16,0,8]]_{\boldsymbol{f}}$ with stabilizer matrix $S_{[[16,0,8]]_{\boldsymbol{f}}}$ $=$ $(\boldsymbol{v}^{(0)}, \boldsymbol{v}^{(1)}, $ $\boldsymbol{v}^{(2)}, \boldsymbol{v}^{(3)},$ $\boldsymbol{v}^{(4)}, \boldsymbol{v}^{(5)}, $ $\boldsymbol{v}^{(1,2)}, \boldsymbol{v}^{(1,3)},$ $ \boldsymbol{v}^{(1,4)}, \boldsymbol{v}^{(1,5)},$ $ \boldsymbol{v}^{(2,3)}, \boldsymbol{v}^{(2,4)},$ $ \boldsymbol{v}^{(2,5)}, \boldsymbol{v}^{(3,4)}, $ $\boldsymbol{v}^{(3,5)}, \boldsymbol{v}^{(4,5)})^{T}$.  More generally, the generator matrix for an RM$(r,m)$ code with $m \ge 2r+1$ corresponds to a fermion code with parameters given in (\ref{eq:RM_Code_f}).  The code distance $d$ is precisely the Hamming distance for the dual code to RM$(r,m)$ code which is the code RM$(m-r-1,m)$.

\section{The 5-qubit code}
The stabilizers for the $[[5,1,3]]$ (five-qubit) code are given by
\begin{align}
\mathcal{O}_{i} \equiv \sigma^{x}_{i}\sigma^{z}_{i+1}\sigma^{z}_{i+2}\sigma^{x}_{i+3}\hspace{.2in} (i = 1,\ldots,4)
\end{align}
with ``periodic boundary conditions", i.e. so that $\hat{\sigma}_{i+5} = \hat{\sigma}_{i}$. All operators commute $[\mathcal{O}_{i},\mathcal{O}_{j}] = 0$ and square to the identity.  The Pauli operators for the encoded qubit are given by
$\hat{Z} \equiv \prod_{i=1}^{5}\sigma^{z}_{i}$, $\hat{X} \equiv \prod_{i=1}^{5}\sigma^{x}_{i}$.  Since the code distance is $d=3$, and the code can correct for single-qubit errors, after performing a non-local unitary (Jordan-Wigner) transformation, local fermion operators that represent {distinct} quasi-particle poisoning events -- which correspond to string operators of the spins -- may have {identical} syndromes.

Performing  a Jordan-Wigner transformation of the 5-qubit code naturally yields a stabilizer code  for Majorana fermions, where the commuting operators now stabilize the fermion parity of the entire system.  Unlike the 5-qubit code, however, not every local operator has a unique syndrome.  The Jordan-Wigner transformation is given by
\begin{align}
\sigma^{z}_{n} &= -i\gamma_{2n-1}\gamma_{2n}\\
\sigma^{x}_{n} &= \left[\prod_{m=1}^{n-1}-i\gamma_{2m-1}\gamma_{2m}\right]\gamma_{2n-1}
\end{align}
where $\gamma_{n}$ are Majorana fermions, satisfying canonical anti-commutation relations $\{\gamma_{n},\gamma_{m}\} = 2\delta_{nm}$.  Observe, for example, that after this transformation
\begin{align}
\gamma_{2} &= \sigma^{y}_{1}\\
\gamma_{7} &= \sigma^{z}_{1}\sigma^{z}_{2}\sigma^{z}_{3}\sigma^{x}_{4}
\end{align}
Within the ground-state subspace of the 5-qubit code, however, both operators have {identical} syndromes since $\gamma_{7}\ket{\Psi_{\mathrm{gs}}} = \gamma_{7}\,\mathcal{O}_{1}\ket{\Psi_{\mathrm{gs}}} = i\gamma_{2}\ket{\Psi_{\mathrm{gs}}}$
so that $\gamma_{7} \sim \gamma_{2}$, up to an overall phase within the ground-state subspace.  More generally, after the Jordan-Wigner transformation
\begin{align}
\mathcal{O}_{1} &= -i\gamma_{2}\gamma_{7}\\
\mathcal{O}_{2} &= -i\gamma_{4}\gamma_{9}\\
\mathcal{O}_{3} &= \gamma_{2}\gamma_{3}\gamma_{4}\gamma_{5}\gamma_{7}\gamma_{8}\gamma_{9}\gamma_{10}\\
\mathcal{O}_{4} &= \gamma_{4}\gamma_{5}\gamma_{6}\gamma_{7}\gamma_{9}\gamma_{10}\gamma_{1}\gamma_{2}\\
\mathcal{O}_{5} &= \gamma_{1}\gamma_{2}\gamma_{3}\gamma_{4}\gamma_{6}\gamma_{7}\gamma_{8}\gamma_{9}
\end{align}
so that the action of the operators $\gamma_{n}$ and $\gamma_{n+5}$ are indistinguishable through measurements of the stabilizer operators:
\begin{align}
\gamma_{n} \sim \gamma_{n+5}
\end{align}


\begin{thebibliography}{1}


\bibitem{Andreev}
A. F. Andreev, JETP {\bf 22}, 455 (1966).

\bibitem{Yu}
L. Yu, Acta Phys. Sin. {\bf 21}, 75 (1965).

\bibitem{Shiba}
H. Shiba, Prog. Theor. Phys. {\bf 40}, 435 (1968).

\bibitem{Rusinov}
A. I. Rusinov, Sov. J. Exp. Theor. Phys. Lett. {\bf 9}, 85 (1969).



\bibitem{Urbina}
C. Janvier {\it et al}, Science, {\bf 349}, 1199 (2015).

\bibitem{BravyiKitaev}
S. Bravyi and A. Kitaev, Annals of Physics, {\bf 298}, 210 (2002).


\bibitem{Kitaev}
A. Kitaev, Phys. Usp. {\bf 44}, 131 (2001).


\bibitem{Read-Green}
N. Read and D. Green, Phys. Rev. B {\bf 61}, 10267 (2000).

\bibitem{Copenhagen1}
M. T. Deng, S. Vaitiekenas, E. B. Hansen, J. Danon, M. Leijnse, K. Flensberg, J. Nygard, P. Krogstrup, C. M. Marcus, Science {\bf 354}, 1557 (2016).

\bibitem{Copenhagen2}  S.  M.  Albrecht,  A.  P.  Higginbotham,  M.  Madsen,  F.
Kuemmeth,  T.  S.  Jespersen,  J.  Nygard,  P.  Krogstrup, and C. M. Marcus, Nature {\bf 531}, 206 (2016).

\bibitem{Delft} V.  Mourik,   K.  Zuo,   S.  M.  Frolov,   S.  R.  Plissard, E. P. A. M. Bakkers,  and L. P. Kouwenhoven,  Science {\bf 336},  1003-1007 (2012).

\bibitem{Princeton} S.  Nadj-Perge,  I.  K.  Drozdov,  J.  Li,  H.  Chen,  S.  Jeon,
J. Seo, A. H. MacDonald, B. A. Bernevig, and A. Yazdani,  Science
{\bf 346}, 602 (2014).

\bibitem{Jia} H.-H. Sun, K.-W. Zhang, L.-H. Hu, C. Li, G.-Y. Wang, H.-Y. Ma, Z.-A. Xu, C.-L. Gao, D.-D. Guan, Y.-Y. Li, C. Liu, D. Qian, Y. Zhou, L. Fu, S.-C. Li, F.-C. Zhang and J.-F. Jia, Phys. Rev. Lett. {\bf 116}, 257003 (2016).

\bibitem{qp1}
P.J. de Visser, D.J. Goldie, P. Diener, S. Withington, J.J.A. Baselmans, T.M. Klapwijk, Phys. Rev. Lett. {\bf 112}, 047004 (2014)

\bibitem{qp2}
M. Zgirski, L. Bretheau, Q. Le Masne, H. Pothier, D. Esteve, and C. Urbina
Phys. Rev. Lett. {\bf 106}, 257003  (2011)

\bibitem{Bravyi}
S. Bravyi, B. Leemhuis, and B. M. Terhal, New J.Phys. {\bf 12}, 083039 (2010).

\bibitem{Vijay}
S. Vijay, T. H. Hsieh, and L. Fu, Phys. Rev. X {\bf 5}, 041038 (2015).

\bibitem{VijaySurface2}
S. Vijay and L. Fu, Phys. Scr. {\bf 2016} 014002

\bibitem{Egger}
L. A. Landau, S. Plugge, E. Sela, A. Altland, S. M. Albrecht, and R. Egger,
Phys. Rev. Lett. {\bf 116}, 050501 (2016).

\bibitem{Egger2}
S. Plugge, L. A. Landau, E. Sela, A. Altland, K. Flensberg, and R. Egger
Phys. Rev. B {\bf 94}, 174514 (2016).

\bibitem{LossSpin}
Silas Hoffman, Constantin Schrade, Jelena Klinovaja, and Daniel Loss
Phys. Rev. B {\bf 94}, 045316  (2016).

\bibitem{Vijay2}
S. Vijay and L. Fu, Phys. Rev. B {\bf 94}, 235446 (2016).

\bibitem{Box}
S. Plugge, A. Rasmussen, R. Egger, K. Flensberg, New J. Phys 19, 012001 (2017).

\bibitem{StationQ}
T. Karzig, C. Knapp, R. Lutchyn, P. Bonderson, M. Hastings, C. Nayak, J. Alicea, K. Flensberg, S. Plugge, Y. Oreg, C. Marcus, M. H. Freedman, arXiv:1610.05289.

\bibitem{FuKane}
 L. Fu and C.L. Kane, Phys. Rev. B {\bf 79}, 161408(R) (2009).

\bibitem{DiVincenzo} F. L. Pedrocchi and D. P. DiVincenzo, Phys. Rev. Lett. {\bf 115}, 120402 (2015).

\bibitem{DiVincenzo_2} F. L. Pedrocchi, N. E. Bonesteel and D. P. DiVincenzo, Phys. Rev. B {\bf 92}, 115441 (2015).

\bibitem{Loss}
D. Rainis and D. Loss, Phys. Rev. B {\bf 85}, 174533 (2012).


\bibitem{Marcus-Parity}
A. P. Higginbotham, S. M. Albrecht, G. Kirsanskas, W. Chang, F. Kuemmeth, P. Krogstrup, T. S. Jespersen, J. Nygard, K. Flensberg and C. M. Marcus, Nature Physics {\bf 11}, 1017 (2015).


\bibitem{Marcus-Parity2}
S. M. Albrecht, E. B. Hansen, A. P. Higginbotham, F. Kuemmeth, T. S. Jespersen, J. Nygard, P. Krogstrup, J. Danon, K. Flensberg, and C. M. Marcus, arXiv:1612.05748



\bibitem{Li}
Y. Li, Phys. Rev. Lett. 117, 120403 (2016)



\bibitem{Gottesman_Thesis} D. Gottesman, \emph{Stabilizer Codes and Quantum Error
Correction}, Ph.D. thesis (CalTech), arXiv:quant-ph/9705052.

\bibitem{Shor_Code} P.W. Shor. Phys. Rev. A., {\bf 52:}R2493, (1995).

\bibitem{Bravyi_Stabilizer_Tradeoff} S. Bravyi, D. Poulin, and B. Terhal, Phys. Rev. Lett. {\bf 104} 050503 (2010).

\bibitem{Bravyi_Kitaev_SC} S. Bravyi and A. Kitaev, arXiv:quant-ph/9811052.

\bibitem{Freedman_SC} M. H. Freedman and D. A. Meyer, Found. Comput. Math. {\bf 1} (3), 325-332.

\bibitem{Fowler_SC} A. G. Fowler, M. Mariantoni, J. M. Martinis, and A. N. Cleland, Phys. Rev. A {\bf 86} (3), 032324.

\bibitem{Kitaev_Spin_Liquid} A. Kitaev, Ann. of Phys. {\bf 321} (2006).

\bibitem{Hamming_Saturate} D. Gottesman. Phys. Rev. A, {\bf 54:}1862-1868, (1996).


\bibitem{Steane_Code}  A.M. Steane, Phys. Rev. Lett., {\bf 77:}793, (1996).

\bibitem{Vijay_Haah_Fu_1} S. Vijay, J. Haah and L. Fu, Phys. Rev. B {\bf 92}, 235136 (2015).

\bibitem{Classical_Coding}  F. J. MacWilliams and N. J. A. Sloane, \emph{The Theory of Error-Correcting Codes}, North-Holland Publishing Company, New York (1977).

\bibitem{CSS} A. R.  Calderbank  and  P. W.  Shor, Phys.  Rev.  A., {\bf 54:}1098, 1996.

\bibitem{Teleportation} L. Fu, Phys. Rev. Lett. {\bf 104}, 056402 2010.



\bibitem{Braiding_without_Braiding} S. Vijay and L. Fu, Phys. Rev. B {\bf 94}, 235446 (2016).











\bibitem{Forthcoming} S. Vijay, unpublished.

\bibitem{SM} Supplemental Material.

\end{thebibliography}
\end{document}